\documentclass[%
reprint,
 amsmath,amssymb,
 aps,
pra,
]{revtex4-1}

\usepackage{graphicx}
\usepackage{dcolumn}
\usepackage{comment}
\usepackage{bm}
\usepackage{placeins} 
\usepackage[colorlinks,citecolor=blue,urlcolor=blue,bookmarks=false,hypertexnames=true]{hyperref} 
\usepackage[mathlines]{lineno}

\begin{document}
\preprint{APS/123-QED}
\title{Optomechanical entanglement at room temperature: a simulation study with realistic conditions}

\author{Kahlil Dixon$^1$}
\author{Lior Cohen$^1$}%
\author{Narayan Bhusal$^1$}
\author{Christopher Wipf$^2$}
\author{Jonathan P. Dowling$^{1,3,4,5}$}
\author{Thomas Corbitt$^1$}
\affiliation{$^1$Department of Physics and Astronomy, Louisiana State University, Baton Rouge, LA 70803, USA}
\affiliation{$^2$LIGO Laboratory, California Institute of Technology, Pasadena, California 91125, USA}
\affiliation{%
	$^3$NYU-ECNU Institute of Physics at NYU Shanghai, 3663 Zhongshan Road North, Shanghai, 200062, China.
	}
\affiliation{%
$^4$CAS-Alibaba Quantum Computing Laboratory, CAS Center for Excellence in Quantum Information and Quantum Physics, University of Science and Technology of China, Shanghai 201315, China.
    }
\affiliation{%
$^5$National Institute of Information and Communications Technology,
4-2-1, Nukui-Kitamachi, Koganei, Tokyo 184-8795, Japan
    }
\date{\today}

\begin{abstract}
Quantum entanglement is the key to many applications like quantum key distribution, quantum teleportation, and quantum sensing. However, reliably generating quantum entanglement in macroscopic systems has proved to be a challenge. Here, we present a detailed analysis of ponderomotive entanglement generation which utilizes optomechanical interactions to create quantum correlations.
We numerically calculate an entanglement measure -- the logarithmic negativity -- for the quantitative assessment of the entanglement. Experimental limitations, including thermal noise and optical loss, from measurements of an existing experiment were included in the calculation, which is intractable to solve analytically.
This work will play an important role in the development of ponderomotive entanglement devices. 
\end{abstract}

\maketitle
\section{\label{sec:level1}Introduction}
Entanglement is the most common and important resource for various quantum technologies, from quantum metrology \cite{giovannetti2011advances, dowling2002quantum}, to quantum communication \cite{ekert1991quantum} and quantum computing \cite{raussendorf2001one,knill2001scheme}. 
It is well known that quantum light sources, in particular entangled sources, require non-linear interaction.
To date, most of these sources are based on all-optical nonlinear processes in crystals \cite{kwiat1995new}, which are good enough for most applications but insufficient for applications with very short wavelengths \cite{sofer2019quantum}.
Recently, efforts have been devoted to explore different avenues to generate quantum entanglement \cite{reiserer2014quantum, PhysRevLett.122.113602, Orieux_2017, alibart2006photon}. One approach is to use strong light-matter interaction with single atoms \cite{reiserer2014quantum} or single quantum dots \cite{PhysRevLett.122.113602}. While this method is very efficient, its production is limited to a single entangled photon pair at a time. A reliable source of entanglement for short wavelength which provides multi-photon entanglement is still in need. 

Radiation pressure \---- the force electromagnetic radiation exerts on a material surface \---- is a significant source of noise in optical metrology \cite{caves1980quantum}. The light's momentum causes fluctuations in the mirror's position, yielding phase noise in the electromagnetic wave. However, this interaction creates quantum correlations that can be exploited to produce non-classical light. It has been shown that when an electromagnetic wave is incident on a mirror, it generates a squeezed light, i.e., the electromagnetic wave experiences an optical nonlinearity \cite{aggarwal2018room}. 
This nonlinearity can also generate entanglement between the light and the mirror \cite{PhysRevLett.98.030405}. Moreover, if two light sources simultaneously interact with the mirror both of them entangle with the mirror, and thus, may entangle with each other \cite{Wipf_2008,wipf_2013,PhysRevLett.110.253601,Barzanjeh2019,Bienfait368}. This form of bipartite optical entanglement generation has been demonstrated experimentally using a vibrating silicon oxide membrane \cite{Chen2020}. This work considers a cantilever $\mu$mirror in-place of the silicon oxide membrane. This oscillating mirror has higher-order modes that should strongly affect entanglement. 



To observe the effect of the quantum back action between the two light fields,  we consider an homodyne quadrature variance measurement of two output optical fields from a single cavity double optical spring with a $\mu$mirror. Here, we report the experimental feasibility of observable ponderomotive entanglement at room temperature and lower temperature. This work identifies experimental configurations that will yield observable entanglement using programs and measurements that have been previously tested and reported ~\cite{PhysRevLett.98.150802,sharifi2019design,cripe_2018}. 
We numerically evaluate the amount of entanglement, and investigate the dependence of entanglement of various experimental parameters such as temperature, side-band frequency, cavity length, and loss.

\section{\label{sec:level2}Methods}
\subsection{ Experimental considerations}
\begin{figure}[ht]
  \centering
 \includegraphics[width=8.6cm]{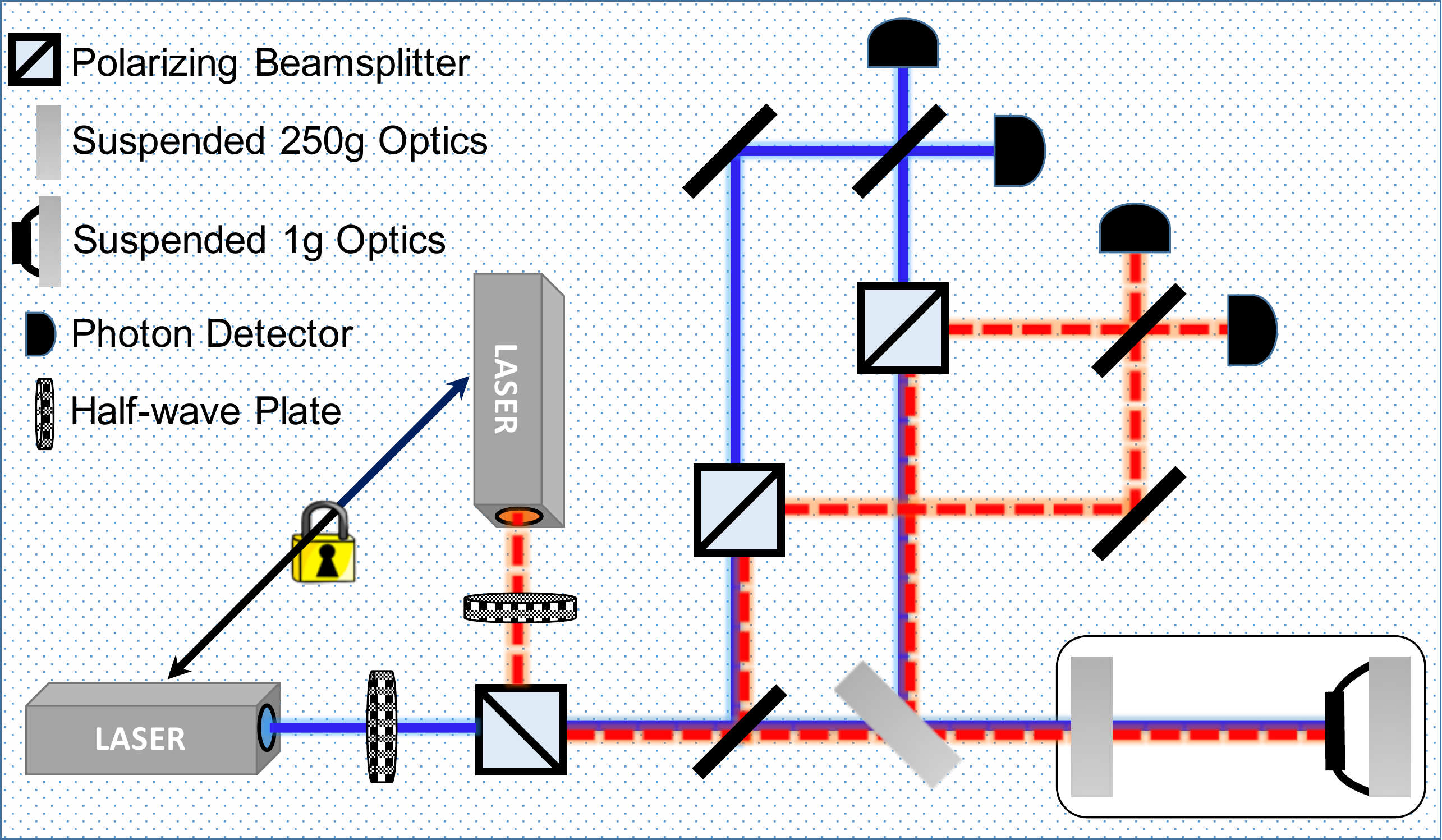}
\caption{Schematic  diagram of the proposed experiment. The scheme uses two frequency locked laser fields that are prepared in two orthogonal polarizations before entering the cavity. A balanced homodyne measurement is employed to construct the full quadrature covariance matrix. The scheme utilizes polarizing beamsplitters to isolate orthogonal polarizations.}
\label{fig:1}
\end{figure}

The experimental consideration is shown in Fig. ~\ref{fig:1}. We chose this configuration because it allows for a stable optomechanical system with no external feedback, which could disrupt the entanglement. It  
utilizes a single optomechanical cavity, that acts as a double optical spring. This setup will convert the squeezing effects of the optical spring cavity into entanglement. Measuring this form of entanglement requires dual homodyne detection to properly measure the squeezing correlations. The two lasers are frequency locked in order to maintain their relative detunings with respect to the cavity resonance. The laser field are arranged with orthogonal relative polarizations.

In order to accurately predict entanglement generation from the $\mu$mirror cavity we take into account: temperature, cavity loss, laser power, and optical spring detunings. Concurrently, we restrict the optical detuning to maintain a stable optical spring; while including more realistic models for input noises. Other variables pertaining to the optomechanical cavity, such as the thermal noise from the $\mu$mirror motion (at room temperature), was taken from experimental data of the same setup~\cite{sharifi2019design}.

\FloatBarrier

\subsection{Measuring entanglement}
To determine whether or not the output fields are entangled we need to choose a convenient measure. The main measure we will use here is the logarithmic negativity entanglement measure ~\cite{PhysRevA.65.032314,PhysRevLett.95.090503}. 
The variance matrix assembled from the quadrature operators will be the main output to measure. The variance matrix can be written as follows:
\begin{eqnarray}
\hspace*{-0.5cm}\mathbf{V}=\left(
\begin{array}{cccc}
 \left\langle  {X_1}^*  {X_1}\right\rangle_+  & \left\langle
    {X_1}^*  {Y_1}\right\rangle_+  & \left\langle
    {X_1}^*  {X_2}\right\rangle_+  & \left\langle
    {X_1}^*  {Y_2}\right\rangle_+  \\
 \left\langle  {Y_1}^*  {X_1}\right\rangle_+  & \left\langle
    {Y_1}^*  {Y_1}\right\rangle_+  & \left\langle
    {Y_1}^*  {X_2}\right\rangle_+  & \left\langle
    {Y_1}^*  {Y_2}\right\rangle_+  \\
 \left\langle  {X_2}^*  {X_1}\right\rangle_+  & \left\langle
    {X_2}^*  {Y_1}\right\rangle_+  & \left\langle
    {X_2}^*  {X_2}\right\rangle_+  & \left\langle
    {X_2}^*  {Y_2}\right\rangle_+  \\
 \left\langle  {Y_2}^*  {X_1}\right\rangle_+  & \left\langle
    {Y_2}^*  {Y_1}\right\rangle_+  & \left\langle
    {Y_2}^*  {X_2}\right\rangle_+  & \left\langle
    {Y_2}^*  {Y_2}\right\rangle_+  \\
\end{array}
\right)
\end{eqnarray} 
were $\left\langle u^*v\right\rangle_+~=~\frac{\left\langle u^*v+ v^*u\right\rangle}{2}$, 
or in block form:
\begin{eqnarray}
\centering
\mathbf{V}= \left(
\begin{array}{cc}
 V_{1 1} & V_{1 2} \\
 V_{2 1} & V_{2 2} \\
\end{array}
\right).
\end{eqnarray}

\subsubsection{Logarithmic negativity}
The logarithmic negativity is useful for measuring continuous-variable (CV) entanglement and is monotone for Gaussian beams (see appendix for an alternate entanglement measure and results). The information-theoretic meaning of logarithmic negativity in terms of exact entanglement cost of quantum Gaussian states was established in~\cite{PhysRevLett.90.027901,wang2018exact,WangWilde}.
Conveniently, the logarithmic negativity can be calculated from the variance matrix~\cite{serafini_2017}:
\begin{eqnarray}
E_N= \max[~0,~ -\ln{\sqrt{2\eta -2\sqrt{\eta^2-4 \det\mathbf{V}}}}~]\\
\textrm{where}~~~ \eta = \det V_{11}+ \det V_{22} -2\det V_{12}.
\end{eqnarray}

\subsection{Computational resources}
While the quantum Langevin approach is more convenient for an analytical approach, to experimentally and computationally develop a simulation, sideband operator propagation is preferred; due to its more intuitive treatment of the optics and higher modularity~\cite{PhysRevA.72.013818,PhysRevA.31.3068}. The simulation assumes an input field and cavity configuration specified by some parameter configuration $\xi$ and outputs the homodyne measurement of the quadratures. It solves for the output quadratures via successive transformation of the input sideband quadratures. To calculate the effect of the micromirror on the input sidebands, measurement data from previous work with the optomechanical cavity is used to simulate the cantilever's effects. These data allow our simulation to consider the cantilever's higher harmonic modes' effect on the entanglement. 

These programs are written to calculate the entanglement measures over different parameter spaces. There are nine adjustable parameters; highlighted in the table \ref{t1}. Past experiments identified configurations that would yield observable single mode squeezing for a single incident beam. These results narrowed our search for optimal parameters. 
\section{\label{sec:level3}Results}

\begin{table}
\centering
$\begin{array}{|c|c|c|}
 \text{Parameter} & \text{Variable
   name} & \text{Stable and $E_N \neq 0$} \\
 \text{Temperature} & T & 295 \text{K} \\
 \text{Circulating carrier } \text{power} & P_1 &
   0.2816 \text{W} \\
 \text{Circulating subcarrier power} &
   P_2 & 0.2238 \text{W} \\
 \text{Loss} & L_s & 250ppm \\
 \text{Carrier } \text{detuning} &
   d_1 & 0.3 \\
 \text{Subcarrier detuning} &
   d_2 & -1.5 \\
 \text{Quality factor} & Q &
   17000 \\
 \text{Cavity } \text{Length} & L_n &
   0.01 \text{m} \\
\end{array}$
\caption{Set of variable simulation parameters, and a stable configuration that yield non-zero entanglement.  }
\label{t1}
\end{table}
The last column in table \ref{t1} represents a parameter set that generates the highest logarithmic negativity and stable optical spring. After the simulations are preformed the output is used to calculate the variance matrices and entanglement measures. All subsequent figures will use the parameters in the table above unless otherwise specified. For example at room temperature and frequency of about 20 kHz, the program predicts an output variance matrix,$\mathbf{V}$ where $E_N(\mathbf{V})=0.104$ and where:
\begin{eqnarray}
\mathbf{V}=\left(
\begin{array}{cccc}
 17.32 &
   -51.38
   &
   -21.06
   &
   -14.80
   \\
 -51.38 &
   156.2 &
   63.76
   & 45.07
   \\
 -21.06 &
   63.76
   &
   26.61
   &
   18.47
   \\
 -14.80 &
   45.07 &
   18.47
   &
   13.54
   \\
\end{array}
\right)
\end{eqnarray}(note that unsqueezed shot noise would have a variance matrix of $\frac{1}{2}\mathbf{I}$ where $\mathbf{I}$ is the identity matrix)  

\begin{figure}[ht]
  \centering
 \includegraphics[width=8.6cm]{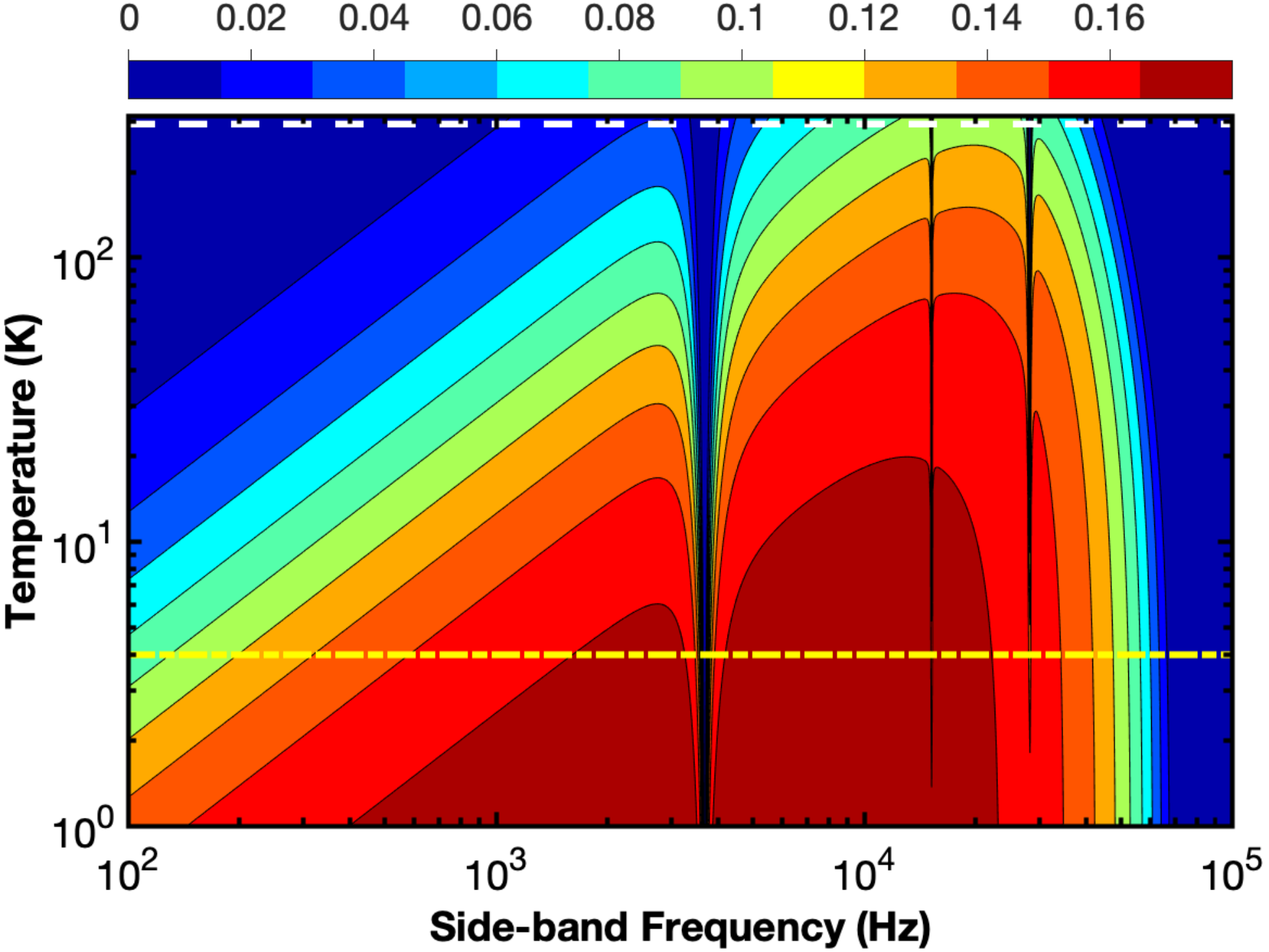}
\caption{Logarithmic negativity measure ($E_N$) of the two output optical fields as a function of temperature and frequency. Conveniently, these parameters yield entanglement at room temperature for a range of frequencies. The sharp drops in $E_N$ are due to the higher order optical spring resonances of the cantilever micromirror (yaw resonance at 4.3 kHz and translation and yaw 54 kHz are the most visible). Top and bottom most dotted lines indicate 295K and 4K respectively.}
\label{fig:2}
\end{figure}

Further analyzing the entanglement yields, we found that the $E_N$ was maximum at 20kHz for the above parameters. This maximum appears to decrease slightly in frequency as temperature decreases as shown in figure \ref{fig:2}. In the figure the sharp drops to zero $E_N$ are at the resonance frequencies of higher order mechanical modes of the cantilever. 

Not only is the double optical spring cavity capable of entangling the two fields, it is able to do so at room temperature. Cooling the micromirror increases both the degree of entanglement and the frequencies over which it is produced. However, there is no significant advantage to cooling the micromirror below 4K, for frequencies of 1kHz and above. This is a result of the thermal noise being pushed well below the quantum back action level, as shown in figure 4. At about 4K, the logarithmic negativity maximizes at approximately $E_N=0.2$ at frequencies above the yaw resonance at 4.3 kHz.  Furthermore, the figure shows that the predicted entanglement closely follows the results of the ponderomotive squeezing experiment, which also maximized at a frequency of about 20 kHz at room temperature~\cite{aggarwal2018room}.

\begin{figure}[ht]
  \centering
 \includegraphics[width=8.6cm]{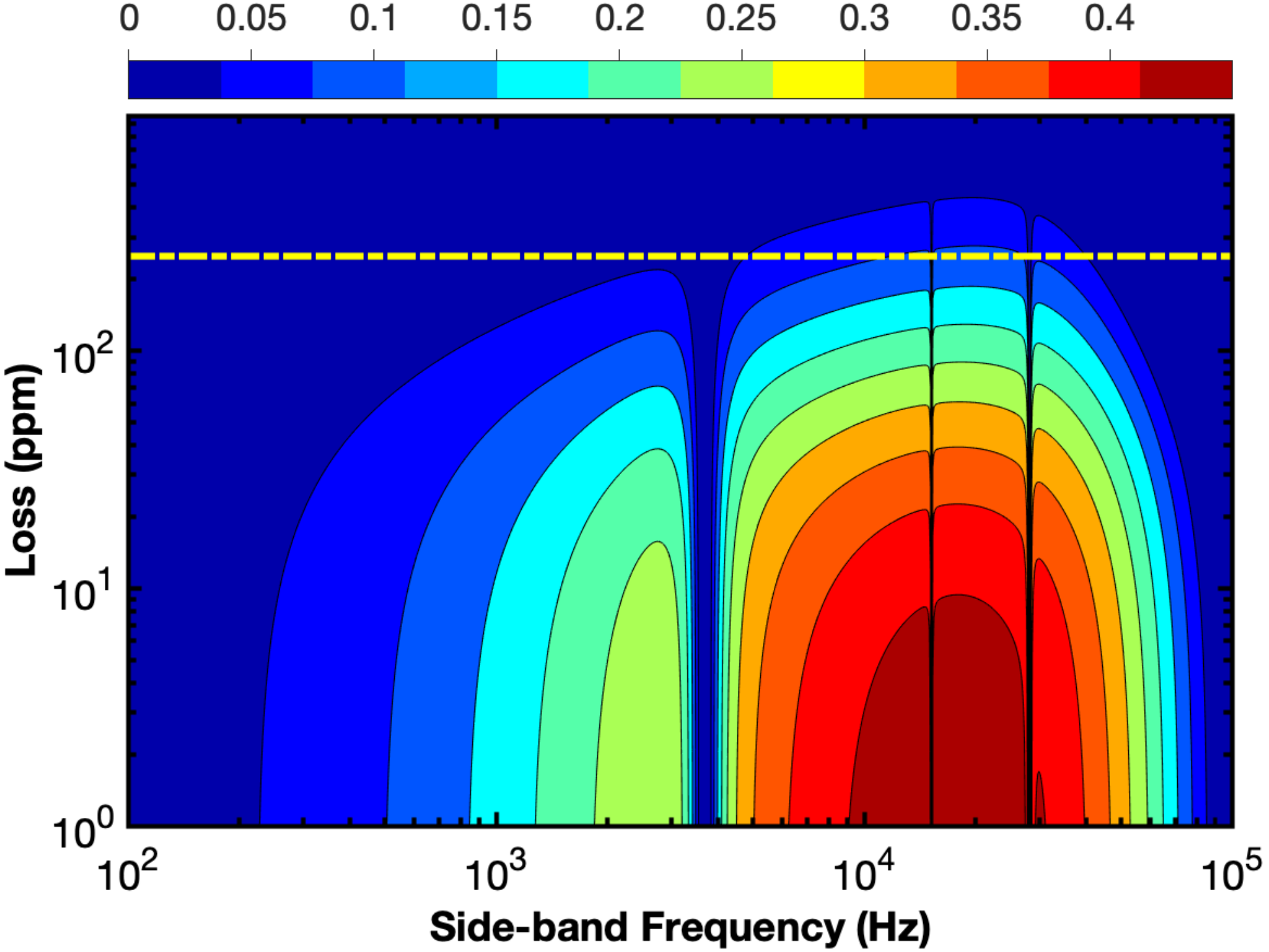}
\caption{$E_N$ vs Loss and frequency at room temperature. Encircled black region denotes absolute zero logarithmic negativity. For small changes in loss the maximum $E_N$ is relatively constant.  The three higher order harmonics are all visible here: yaw,ya-transverse, roll-transverse. The yellow line indicates current experimental losses. }
\label{fig:3}
\end{figure}

More promise for this method is inspired by the results displayed in figure \ref{fig:3}. Even when realistic noise and losses are considered the entanglement persists. 

Lower losses also aid the entangler; figure \ref{fig:3} shows the entanglement increases as loss decreases. The behavior of the classical to quantum noise ratio well follows that of the entanglement at all temperatures.

\begin{figure}[ht]
  \centering
 \includegraphics[width=8.6cm]{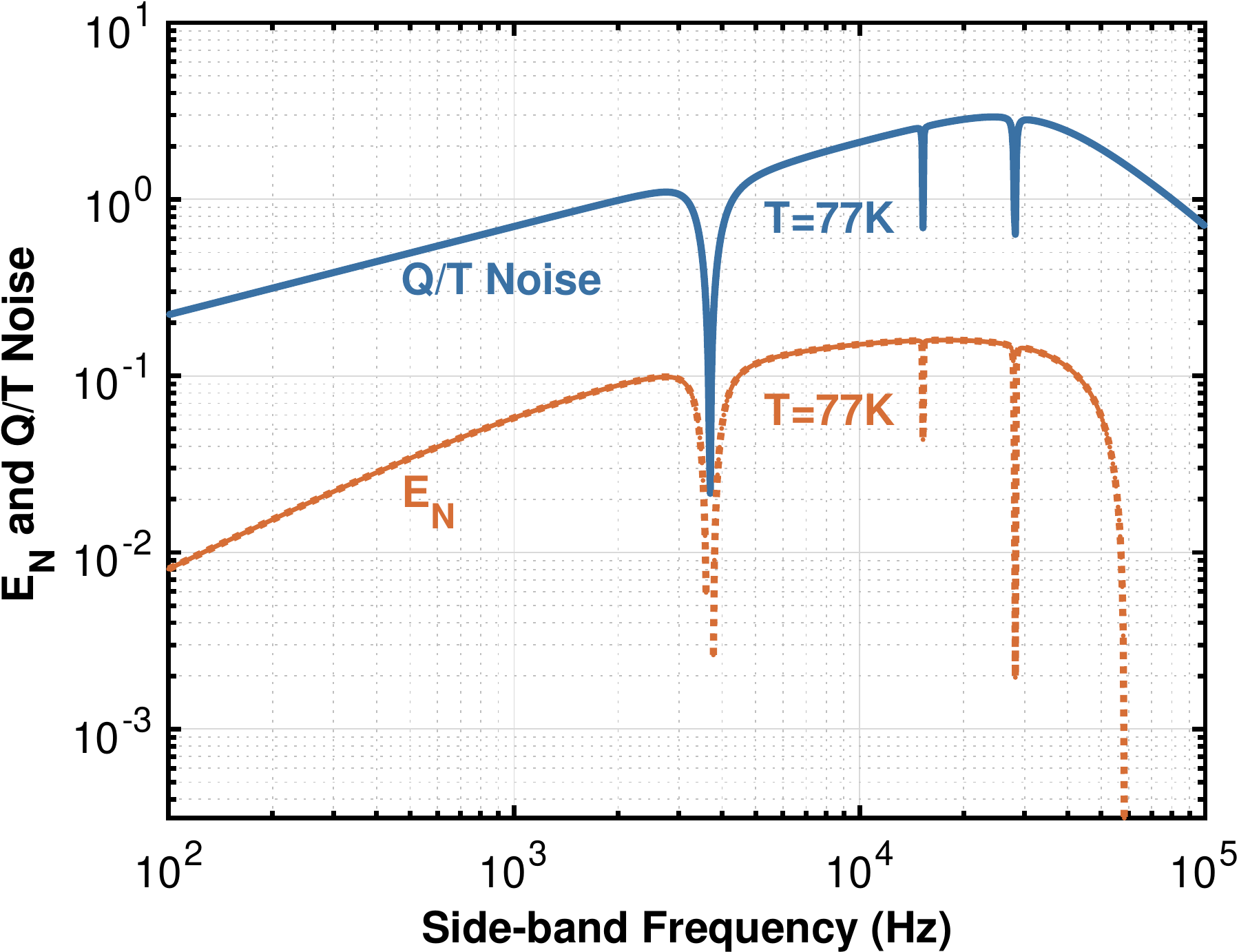}
\caption{The ratio of quantum to thermal noises in the system. The Logarithmic negativity result at 77K has been included to show agreement. While this confirms our hypothesis about the quantum radiation pressure noise working in opposition of the thermal noises to yield entanglement, in conjunction with our other results it also shows that such an entanglement generation technique heavily relies on the other experimental parameters as well.}
\label{fig:4}
\end{figure}
\begin{figure}[ht]
  \centering
 \includegraphics[width=8.6cm]{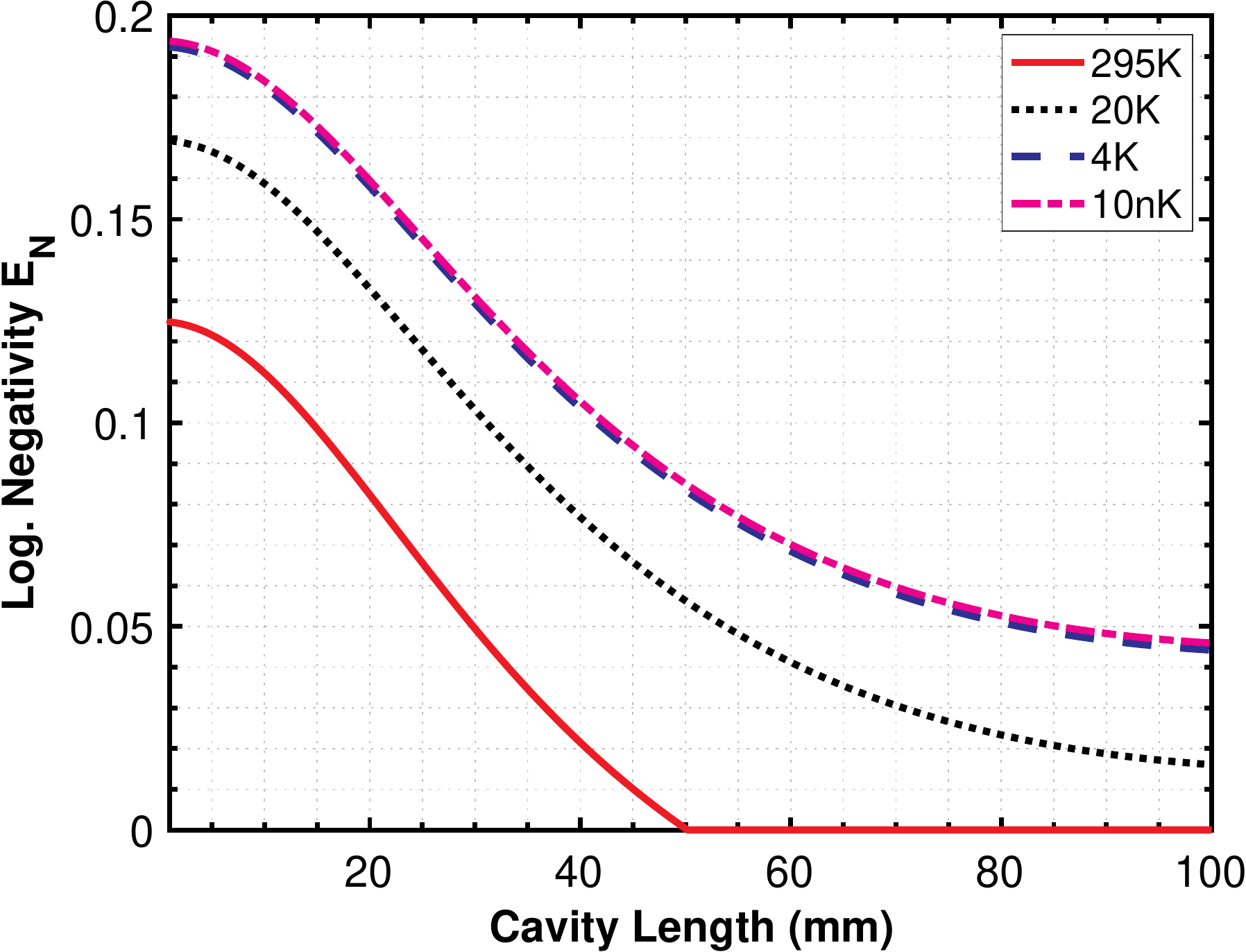}
\caption{Logarithmic negativity versus the cavity length for different ambient temperatures at 20kHz. No benefit to entanglement generation will be seen when cooling below 4K (unless operating at the cantilever's higher harmonic frequencies). Furthermore, cooling below 1K avoids losses to entanglement due to harmonic effects the cavity length effects on the entanglement (the drop at 10cm) are present at all temperatures. Entanglement does not improve at cavity lengths shorter than 1mm. }
\label{fig:5}
\end{figure}

To maximize entanglement we consider changing the optomechanical cavity length. Figure \ref{fig:5} shows the dependence between cavity length and $E_N$.  

Together figures ~\ref{fig:4} and ~\ref{fig:5} pertain to the fundamental concepts behind optomechanical entanglement generation. The cavity length changes due to the input laser power. This has quantum fluctuations due to the  Heisenberg uncertainty principle. This creates a fundamental uncertainty in the overall cavity length, which in turn strongly effects the properties of the output light. This technique manipulates quantum radiation pressure noise into an entanglement source. When this noise is greater than the classical noise, in this case thermal noise, the entanglement should thrive; figure ~\ref{fig:4} confirms this. Furthermore, entanglement will be limited if the cavity length fluctuations are too small relative to the overall cavity length. Moreover, the dampening effects become more dominant as the cavity length increases thus widening the resonances that destroy entanglement.  This is shown in figure ~\ref{fig:5}.  

\begin{figure}[ht]
  \centering
 \includegraphics[width=8.6cm]{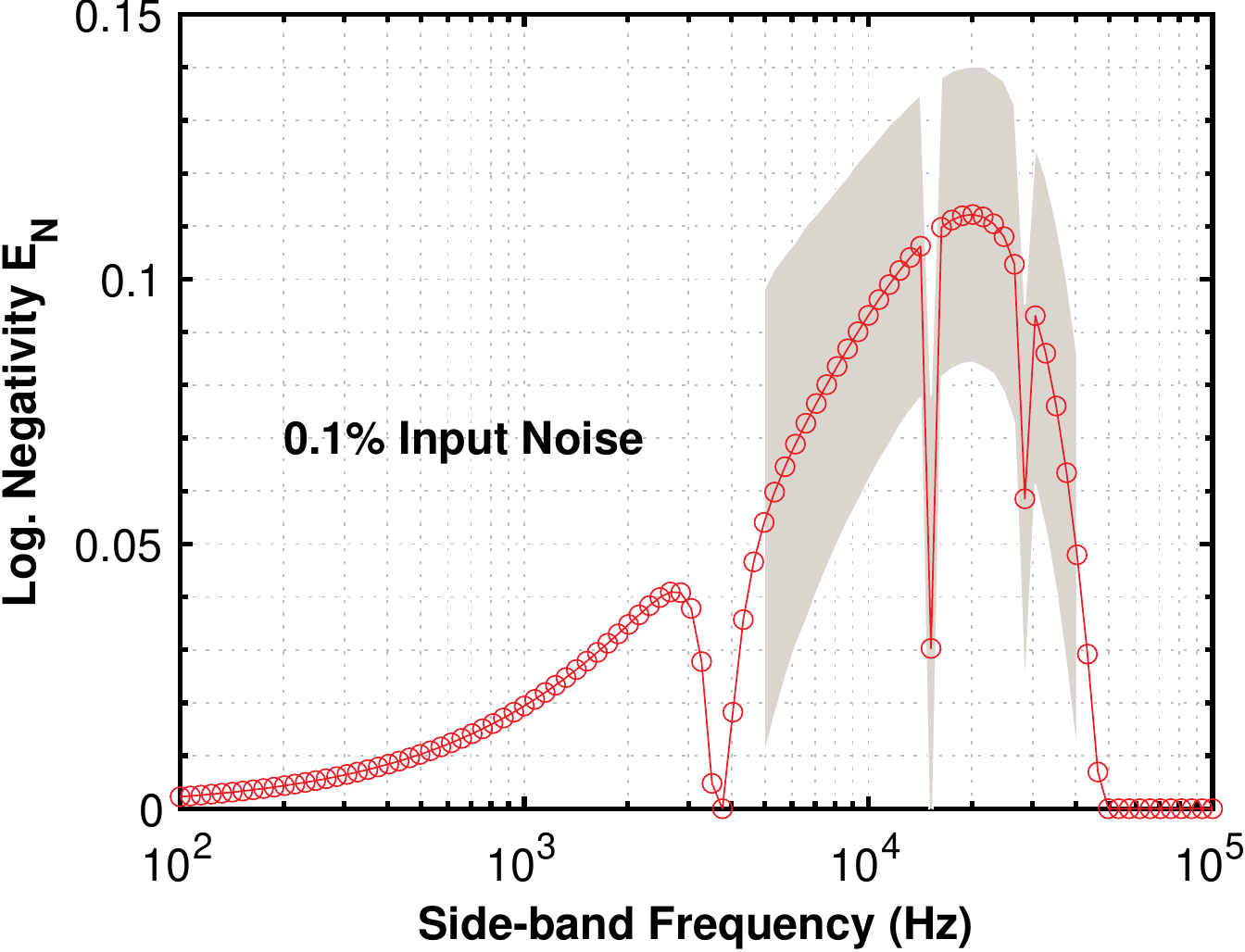}
\caption{We can simulate the effects of Gaussian noise in the double homodyne measurement by creating a normal distribution of variance matrices and calculating the standard deviation in the resulting negativity.  Due to the numeric instabilities in the entanglement measure, entanglement verification requires high precision measurement of the output quadratures (about 0.1\%) (67\% confidence interval) ; the logarithmic negativity is highly sensitive to Gaussian noise. The grey shaded region denotes the uncertainty in the output entanglement. All parameters for this calculation match those in the table previous.}
\label{gn1}
\end{figure}

We would like to quantify how difficult it is to experimentally verify the existence of the simulated entanglement. We simulate a noisy variance matrix measurement by creating a set of variance matrices normally distributed about the initial output variance matrix at each frequency. (While we shall only show the experiments noise sensitivity as a function of Gaussian spread and frequency, it is possible to vary any of the parameters in the table for the noise analysis.) The resulting entanglement uncertainties are plotted in figure ~\ref{gn1}.

With measurement certainty on the order of 1\% the output noise in the measurement is several times that of the expected maximum entanglement. When measuring at or near the peak $E_N$ frequency, the double homodyne precision must be on the order of 0.1\%. 

\FloatBarrier
\section{\label{sec:level4}Conclusion}

All optical circuits and devices are subject to quantum radiation pressure effects. These effects correlate incident light; which implies potential for new entanglement devices. The effects are strong enough to be manipulated into generating bipartite optical entanglement. Moreover, this entanglement persists at room temperature with realistic losses, stable optical spring detunings, and accessible circulating powers. With experimentally stable parameters, we predict a maximum logarithmic negativity of $E_N=0.2$; while considering parameters close to reported experiments yields average logarithmic negativity of $E_N= 0.3$ (with about 1\% measurement certainty) which agrees with the results reported there \cite{Chen2020}.  Furthermore, we found that entanglement is highly temperature dependent.  While lowering losses could enhance entanglement, we have shown that the current loss levels still allow for entanglement. Although, predicted entanglement persists despite realistic noise and higher mode  considerations, the sensitivity of the system to Gaussian noises presents a significant challenge to experimental realizations.  Further optimization may be required to achieve accessible entanglement output. 

\section*{Acknowledgement}
K.D., L.C., N.B. and J.P.D. would like to acknowledge the Air Force Office of Scientific Research, the Army Research Office, the Defense Advanced Research Projects Agency, and the National Science Foundation. This material is based upon work supported by the National Science Foundation under Grant No. PHY-1806634. We would like to thank X,Y,Z for important discussions.

\section{\label{sec:level5} Appendix}
\subsection{Logarithmic negativity} 
Negativity is an "easy-to-compute" measure of entanglement defined as follows:
\begin{eqnarray}\
\centering
N(\rho) = \frac{||\rho^{\Gamma_{A}}||-1}{2}
\end{eqnarray}
,where $\rho$ is the density matrix, $A$ is the dimension of the subsystem, and $\rho^{\Gamma_A}$ is the partial transpose of $\rho$ with respect to subsystem $A$~\cite{PhysRevA.60.179,buml2019resource}. Written with the same dependence the logarithmic negativity is the following:
\begin{eqnarray}\
\centering
E_N = log_2||\rho^{\Gamma_A}||_1.
\end{eqnarray}
\subsection{Duan's measure of inseparability}
Since the logarithmic negativity is strongly dependent on our normalization we compute a second entanglement measure as a sanity check. We chose this measure because it does not vary with choice of variance matrix normalization. This entanglement monotone is an alternative to the negativity based measure for CV entangled beams~\cite{PhysRevLett.84.2722}. 
The calculation/ determining of the "$a$" parameter is dependent on calculations done on/with the variance matrix "$V$"; however, the only variance matrices of certain forms can be used~\cite{PhysRevLett.84.2722}. Fortunately, Duan proved that non standard form variance matrices can be transformed into their standard forms following a few steps and solving a few equations.  The variance matrix of the standard form (the goal) shall be written as follows:
\begin{eqnarray}\
\centering
V'' =\left(
\begin{array}{cccc}
 n_1 &  & c_1 &  \\
  & n_2 &  & c_2 \\
 c_1 &  & m_1 &  \\
  & c_2 &  & m_2 \\
\end{array}
\right).
\end{eqnarray}
These matrix elements are computed from the elements of what we shall call the "substandard form of the variance matrix"; this form shall be written as follows:
\begin{eqnarray}\
\centering
V' =\left(
\begin{array}{cccc}
 n &  & c &  \\
  & n &  & c' \\
 c &  & m &  \\
  & c' &  & m \\
\end{array}
\right).
\end{eqnarray}
The standard form is calculated from the substandard form by solving the following system of equations for the parameters $r_1$ and $r_2$:
\begin{eqnarray}
\sqrt{r_1r_2}|c|-\frac{|c'|}{\sqrt{r_1r_2}}=\sqrt{\alpha_n \alpha_m}-\sqrt{(\beta_n\beta_m)}\\
\frac{\beta_n}{\alpha_n}=\frac{\beta_m}{\alpha_m}
\end{eqnarray}
where $\alpha_n=nr_1-1$, $\beta_n= \frac{n}{r_1}-1$, $\alpha_m=mr_2-1$, and $\beta_m=\frac{m}{r_2}-1$. Then, to apply $r_1$ and $r_2$: $n_1=nr_1,~n_2=n/r_2,~m_1=mr_1,~m_2=mr_2,~c_1=c\sqrt{r_1r_2},~c_2=\frac{c'}{\sqrt{r_1r_2}}$.
Finally, our original variance matrix "$V$" will be reference in block form as follows:
\begin{eqnarray}\
\centering
V =\left(
\begin{array}{cc}
 V_{11} &  V_{12}   \\
 V _{12}^T  & V_{22} \\
\end{array}
\right) =\left(
\begin{array}{cc}
 A&B  \\
  B^T &C \\
\end{array}
\right).
\end{eqnarray}.
Calculating $\mathbf{V}'$ from $\mathbf{V}$ was done using the following equations. 
\begin{eqnarray}
\det A = n^2,~~ \det C = m^2, \det B = c c',\\
\det \mathbf{V} = (nm-c^2)(nm-c'^2)
\end{eqnarray}

After attaining the proper form, the following inequalities need to be broken for there to be entanglement in the system:
\begin{eqnarray}
|c_1| \leq \sqrt{(n_1-1) (m_1-1)}\\
|c_2| \leq \sqrt{(n_2-1) (m_2-1)}
\end{eqnarray}
To compare system inseparabilities we use another form of the metric inequality:
\begin{eqnarray}
R = \frac{\frac{a^2(n_1+n_2)}{2}+\frac{(m_1+m_2)}{2a^2}-|c_1|-|c_2|}{a^2+a^{-2}}\geq 1\\
\textrm{where}~ a^2=\sqrt{{\frac{m_1-1}{n_1-1}}},
\end{eqnarray}
and the system is only separable when this inequality is satisfied. 

\begin{figure}[ht]
  \centering
 \includegraphics[width=8.6cm]{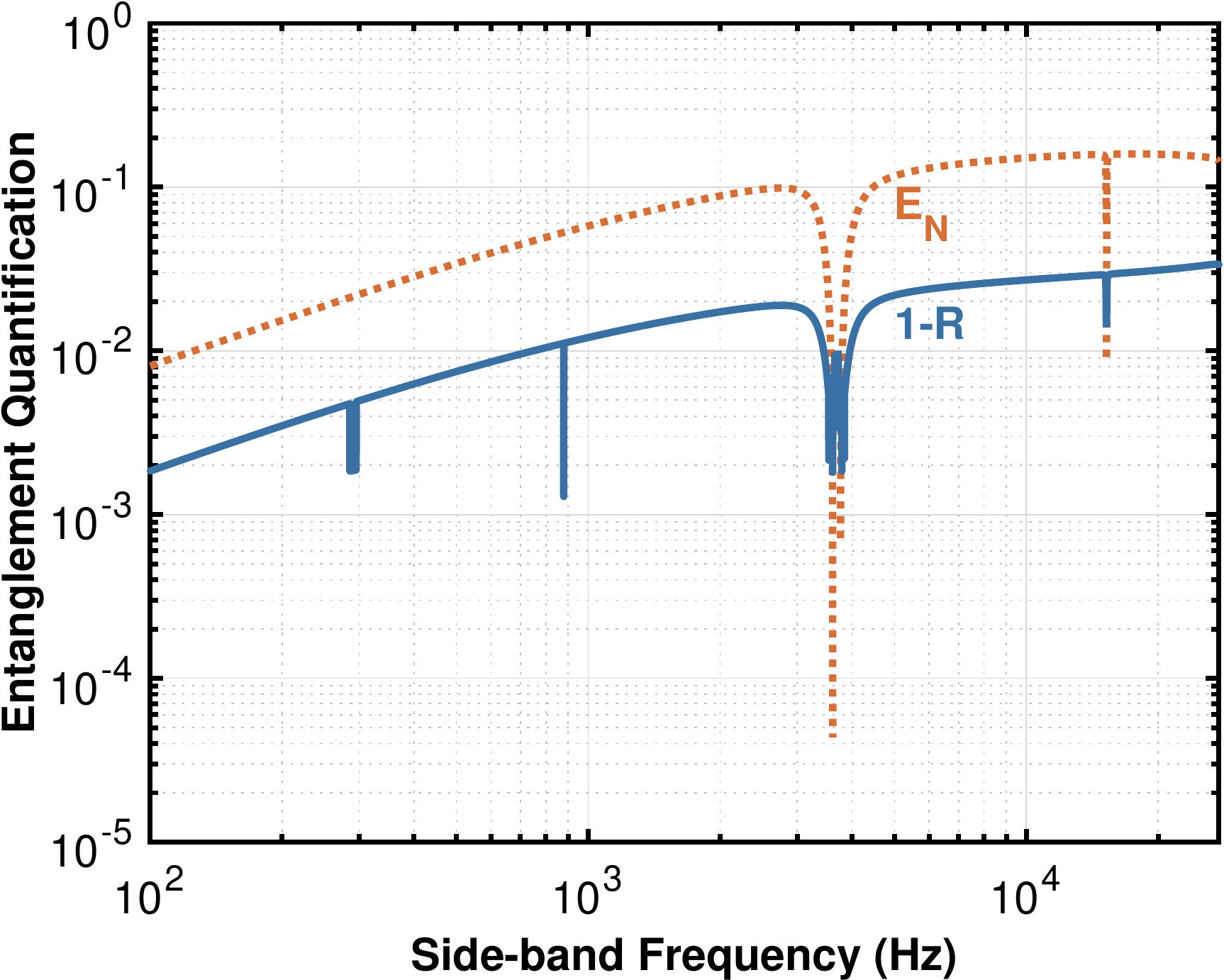}
\caption{Duan measure $1-R$ and logarithmic negativity $E_N$ versus side-band frequency.For this measure, higher values correspond to stronger entanglement. Both at T = 295 K.}
\label{fig:6}
\end{figure}


\FloatBarrier
\bibliography{mybib}

\begin{thebibliography}{34}%
\makeatletter
\providecommand \@ifxundefined [1]{%
 \@ifx{#1\undefined}
}%
\providecommand \@ifnum [1]{%
 \ifnum #1\expandafter \@firstoftwo
 \else \expandafter \@secondoftwo
 \fi
}%
\providecommand \@ifx [1]{%
 \ifx #1\expandafter \@firstoftwo
 \else \expandafter \@secondoftwo
 \fi
}%
\providecommand \natexlab [1]{#1}%
\providecommand \enquote  [1]{``#1''}%
\providecommand \bibnamefont  [1]{#1}%
\providecommand \bibfnamefont [1]{#1}%
\providecommand \citenamefont [1]{#1}%
\providecommand \href@noop [0]{\@secondoftwo}%
\providecommand \href [0]{\begingroup \@sanitize@url \@href}%
\providecommand \@href[1]{\@@startlink{#1}\@@href}%
\providecommand \@@href[1]{\endgroup#1\@@endlink}%
\providecommand \@sanitize@url [0]{\catcode `\\12\catcode `\$12\catcode
  `\&12\catcode `\#12\catcode `\^12\catcode `\_12\catcode `\%12\relax}%
\providecommand \@@startlink[1]{}%
\providecommand \@@endlink[0]{}%
\providecommand \url  [0]{\begingroup\@sanitize@url \@url }%
\providecommand \@url [1]{\endgroup\@href {#1}{\urlprefix }}%
\providecommand \urlprefix  [0]{URL }%
\providecommand \Eprint [0]{\href }%
\providecommand \doibase [0]{http://dx.doi.org/}%
\providecommand \selectlanguage [0]{\@gobble}%
\providecommand \bibinfo  [0]{\@secondoftwo}%
\providecommand \bibfield  [0]{\@secondoftwo}%
\providecommand \translation [1]{[#1]}%
\providecommand \BibitemOpen [0]{}%
\providecommand \bibitemStop [0]{}%
\providecommand \bibitemNoStop [0]{.\EOS\space}%
\providecommand \EOS [0]{\spacefactor3000\relax}%
\providecommand \BibitemShut  [1]{\csname bibitem#1\endcsname}%
\let\auto@bib@innerbib\@empty
\bibitem [{\citenamefont {Giovannetti}\ \emph {et~al.}(2011)\citenamefont
  {Giovannetti}, \citenamefont {Lloyd},\ and\ \citenamefont
  {Maccone}}]{giovannetti2011advances}%
  \BibitemOpen
  \bibfield  {author} {\bibinfo {author} {\bibfnamefont {V.}~\bibnamefont
  {Giovannetti}}, \bibinfo {author} {\bibfnamefont {S.}~\bibnamefont {Lloyd}},
  \ and\ \bibinfo {author} {\bibfnamefont {L.}~\bibnamefont {Maccone}},\
  }\href@noop {} {\bibfield  {journal} {\bibinfo  {journal} {Nature photonics}\
  }\textbf {\bibinfo {volume} {5}},\ \bibinfo {pages} {222} (\bibinfo {year}
  {2011})}\BibitemShut {NoStop}%
\bibitem [{\citenamefont {Dowling}\ and\ \citenamefont
  {Milburn}(2002)}]{dowling2002quantum}%
  \BibitemOpen
  \bibfield  {author} {\bibinfo {author} {\bibfnamefont {J.~P.}\ \bibnamefont
  {Dowling}}\ and\ \bibinfo {author} {\bibfnamefont {G.~J.}\ \bibnamefont
  {Milburn}},\ }\href@noop {} {\enquote {\bibinfo {title} {Quantum technology:
  The second quantum revolution},}\ } (\bibinfo {year} {2002}),\ \Eprint
  {http://arxiv.org/abs/quant-ph/0206091} {arXiv:quant-ph/0206091 [quant-ph]}
  \BibitemShut {NoStop}%
\bibitem [{\citenamefont {Ekert}(1991)}]{ekert1991quantum}%
  \BibitemOpen
  \bibfield  {author} {\bibinfo {author} {\bibfnamefont {A.~K.}\ \bibnamefont
  {Ekert}},\ }\href@noop {} {\bibfield  {journal} {\bibinfo  {journal}
  {Physical review letters}\ }\textbf {\bibinfo {volume} {67}},\ \bibinfo
  {pages} {661} (\bibinfo {year} {1991})}\BibitemShut {NoStop}%
\bibitem [{\citenamefont {Raussendorf}\ and\ \citenamefont
  {Briegel}(2001)}]{raussendorf2001one}%
  \BibitemOpen
  \bibfield  {author} {\bibinfo {author} {\bibfnamefont {R.}~\bibnamefont
  {Raussendorf}}\ and\ \bibinfo {author} {\bibfnamefont {H.~J.}\ \bibnamefont
  {Briegel}},\ }\href@noop {} {\bibfield  {journal} {\bibinfo  {journal}
  {Physical Review Letters}\ }\textbf {\bibinfo {volume} {86}},\ \bibinfo
  {pages} {5188} (\bibinfo {year} {2001})}\BibitemShut {NoStop}%
\bibitem [{\citenamefont {Knill}\ \emph {et~al.}(2001)\citenamefont {Knill},
  \citenamefont {Laflamme},\ and\ \citenamefont {Milburn}}]{knill2001scheme}%
  \BibitemOpen
  \bibfield  {author} {\bibinfo {author} {\bibfnamefont {E.}~\bibnamefont
  {Knill}}, \bibinfo {author} {\bibfnamefont {R.}~\bibnamefont {Laflamme}}, \
  and\ \bibinfo {author} {\bibfnamefont {G.~J.}\ \bibnamefont {Milburn}},\
  }\href@noop {} {\bibfield  {journal} {\bibinfo  {journal} {nature}\ }\textbf
  {\bibinfo {volume} {409}},\ \bibinfo {pages} {46} (\bibinfo {year}
  {2001})}\BibitemShut {NoStop}%
\bibitem [{\citenamefont {Kwiat}\ \emph {et~al.}(1995)\citenamefont {Kwiat},
  \citenamefont {Mattle}, \citenamefont {Weinfurter}, \citenamefont
  {Zeilinger}, \citenamefont {Sergienko},\ and\ \citenamefont
  {Shih}}]{kwiat1995new}%
  \BibitemOpen
  \bibfield  {author} {\bibinfo {author} {\bibfnamefont {P.~G.}\ \bibnamefont
  {Kwiat}}, \bibinfo {author} {\bibfnamefont {K.}~\bibnamefont {Mattle}},
  \bibinfo {author} {\bibfnamefont {H.}~\bibnamefont {Weinfurter}}, \bibinfo
  {author} {\bibfnamefont {A.}~\bibnamefont {Zeilinger}}, \bibinfo {author}
  {\bibfnamefont {A.~V.}\ \bibnamefont {Sergienko}}, \ and\ \bibinfo {author}
  {\bibfnamefont {Y.}~\bibnamefont {Shih}},\ }\href@noop {} {\bibfield
  {journal} {\bibinfo  {journal} {Physical Review Letters}\ }\textbf {\bibinfo
  {volume} {75}},\ \bibinfo {pages} {4337} (\bibinfo {year}
  {1995})}\BibitemShut {NoStop}%
\bibitem [{\citenamefont {Sofer}\ \emph {et~al.}(2019)\citenamefont {Sofer},
  \citenamefont {Strizhevsky}, \citenamefont {Schori}, \citenamefont
  {Tamasaku},\ and\ \citenamefont {Shwartz}}]{sofer2019quantum}%
  \BibitemOpen
  \bibfield  {author} {\bibinfo {author} {\bibfnamefont {S.}~\bibnamefont
  {Sofer}}, \bibinfo {author} {\bibfnamefont {E.}~\bibnamefont {Strizhevsky}},
  \bibinfo {author} {\bibfnamefont {A.}~\bibnamefont {Schori}}, \bibinfo
  {author} {\bibfnamefont {K.}~\bibnamefont {Tamasaku}}, \ and\ \bibinfo
  {author} {\bibfnamefont {S.}~\bibnamefont {Shwartz}},\ }\href@noop {}
  {\bibfield  {journal} {\bibinfo  {journal} {Physical Review X}\ }\textbf
  {\bibinfo {volume} {9}},\ \bibinfo {pages} {031033} (\bibinfo {year}
  {2019})}\BibitemShut {NoStop}%
\bibitem [{\citenamefont {Reiserer}\ \emph {et~al.}(2014)\citenamefont
  {Reiserer}, \citenamefont {Kalb}, \citenamefont {Rempe},\ and\ \citenamefont
  {Ritter}}]{reiserer2014quantum}%
  \BibitemOpen
  \bibfield  {author} {\bibinfo {author} {\bibfnamefont {A.}~\bibnamefont
  {Reiserer}}, \bibinfo {author} {\bibfnamefont {N.}~\bibnamefont {Kalb}},
  \bibinfo {author} {\bibfnamefont {G.}~\bibnamefont {Rempe}}, \ and\ \bibinfo
  {author} {\bibfnamefont {S.}~\bibnamefont {Ritter}},\ }\href@noop {}
  {\bibfield  {journal} {\bibinfo  {journal} {Nature}\ }\textbf {\bibinfo
  {volume} {508}},\ \bibinfo {pages} {237} (\bibinfo {year}
  {2014})}\BibitemShut {NoStop}%
\bibitem [{\citenamefont {Wang}\ \emph {et~al.}(2019)\citenamefont {Wang},
  \citenamefont {Hu}, \citenamefont {Chung}, \citenamefont {Qin}, \citenamefont
  {Yang}, \citenamefont {Li}, \citenamefont {Liu}, \citenamefont {Zhong},
  \citenamefont {He}, \citenamefont {Ding}, \citenamefont {Deng}, \citenamefont
  {Dai}, \citenamefont {Huo}, \citenamefont {H\"ofling}, \citenamefont {Lu},\
  and\ \citenamefont {Pan}}]{PhysRevLett.122.113602}%
  \BibitemOpen
  \bibfield  {author} {\bibinfo {author} {\bibfnamefont {H.}~\bibnamefont
  {Wang}}, \bibinfo {author} {\bibfnamefont {H.}~\bibnamefont {Hu}}, \bibinfo
  {author} {\bibfnamefont {T.-H.}\ \bibnamefont {Chung}}, \bibinfo {author}
  {\bibfnamefont {J.}~\bibnamefont {Qin}}, \bibinfo {author} {\bibfnamefont
  {X.}~\bibnamefont {Yang}}, \bibinfo {author} {\bibfnamefont {J.-P.}\
  \bibnamefont {Li}}, \bibinfo {author} {\bibfnamefont {R.-Z.}\ \bibnamefont
  {Liu}}, \bibinfo {author} {\bibfnamefont {H.-S.}\ \bibnamefont {Zhong}},
  \bibinfo {author} {\bibfnamefont {Y.-M.}\ \bibnamefont {He}}, \bibinfo
  {author} {\bibfnamefont {X.}~\bibnamefont {Ding}}, \bibinfo {author}
  {\bibfnamefont {Y.-H.}\ \bibnamefont {Deng}}, \bibinfo {author}
  {\bibfnamefont {Q.}~\bibnamefont {Dai}}, \bibinfo {author} {\bibfnamefont
  {Y.-H.}\ \bibnamefont {Huo}}, \bibinfo {author} {\bibfnamefont
  {S.}~\bibnamefont {H\"ofling}}, \bibinfo {author} {\bibfnamefont {C.-Y.}\
  \bibnamefont {Lu}}, \ and\ \bibinfo {author} {\bibfnamefont {J.-W.}\
  \bibnamefont {Pan}},\ }\href {\doibase 10.1103/PhysRevLett.122.113602}
  {\bibfield  {journal} {\bibinfo  {journal} {Phys. Rev. Lett.}\ }\textbf
  {\bibinfo {volume} {122}},\ \bibinfo {pages} {113602} (\bibinfo {year}
  {2019})}\BibitemShut {NoStop}%
\bibitem [{\citenamefont {Orieux}\ \emph {et~al.}(2017)\citenamefont {Orieux},
  \citenamefont {Versteegh}, \citenamefont {Jöns},\ and\ \citenamefont
  {Ducci}}]{Orieux_2017}%
  \BibitemOpen
  \bibfield  {author} {\bibinfo {author} {\bibfnamefont {A.}~\bibnamefont
  {Orieux}}, \bibinfo {author} {\bibfnamefont {M.~A.~M.}\ \bibnamefont
  {Versteegh}}, \bibinfo {author} {\bibfnamefont {K.~D.}\ \bibnamefont
  {Jöns}}, \ and\ \bibinfo {author} {\bibfnamefont {S.}~\bibnamefont
  {Ducci}},\ }\href {\doibase 10.1088/1361-6633/aa6955} {\bibfield  {journal}
  {\bibinfo  {journal} {Reports on Progress in Physics}\ }\textbf {\bibinfo
  {volume} {80}},\ \bibinfo {pages} {076001} (\bibinfo {year}
  {2017})}\BibitemShut {NoStop}%
\bibitem [{\citenamefont {Alibart}\ \emph {et~al.}(2006)\citenamefont
  {Alibart}, \citenamefont {Fulconis}, \citenamefont {Wong}, \citenamefont
  {Murdoch}, \citenamefont {Wadsworth},\ and\ \citenamefont
  {Rarity}}]{alibart2006photon}%
  \BibitemOpen
  \bibfield  {author} {\bibinfo {author} {\bibfnamefont {O.}~\bibnamefont
  {Alibart}}, \bibinfo {author} {\bibfnamefont {J.}~\bibnamefont {Fulconis}},
  \bibinfo {author} {\bibfnamefont {G.}~\bibnamefont {Wong}}, \bibinfo {author}
  {\bibfnamefont {S.}~\bibnamefont {Murdoch}}, \bibinfo {author} {\bibfnamefont
  {W.}~\bibnamefont {Wadsworth}}, \ and\ \bibinfo {author} {\bibfnamefont
  {J.}~\bibnamefont {Rarity}},\ }\href@noop {} {\bibfield  {journal} {\bibinfo
  {journal} {New Journal of Physics}\ }\textbf {\bibinfo {volume} {8}},\
  \bibinfo {pages} {67} (\bibinfo {year} {2006})}\BibitemShut {NoStop}%
\bibitem [{\citenamefont {Caves}(1980)}]{caves1980quantum}%
  \BibitemOpen
  \bibfield  {author} {\bibinfo {author} {\bibfnamefont {C.~M.}\ \bibnamefont
  {Caves}},\ }\href@noop {} {\bibfield  {journal} {\bibinfo  {journal}
  {Physical Review Letters}\ }\textbf {\bibinfo {volume} {45}},\ \bibinfo
  {pages} {75} (\bibinfo {year} {1980})}\BibitemShut {NoStop}%
\bibitem [{\citenamefont {Aggarwal}\ \emph {et~al.}(2018)\citenamefont
  {Aggarwal}, \citenamefont {Cullen}, \citenamefont {Cripe}, \citenamefont
  {Cole}, \citenamefont {Lanza}, \citenamefont {Libson}, \citenamefont
  {Follman}, \citenamefont {Heu}, \citenamefont {Corbitt},\ and\ \citenamefont
  {Mavalvala}}]{aggarwal2018room}%
  \BibitemOpen
  \bibfield  {author} {\bibinfo {author} {\bibfnamefont {N.}~\bibnamefont
  {Aggarwal}}, \bibinfo {author} {\bibfnamefont {T.}~\bibnamefont {Cullen}},
  \bibinfo {author} {\bibfnamefont {J.}~\bibnamefont {Cripe}}, \bibinfo
  {author} {\bibfnamefont {G.~D.}\ \bibnamefont {Cole}}, \bibinfo {author}
  {\bibfnamefont {R.}~\bibnamefont {Lanza}}, \bibinfo {author} {\bibfnamefont
  {A.}~\bibnamefont {Libson}}, \bibinfo {author} {\bibfnamefont
  {D.}~\bibnamefont {Follman}}, \bibinfo {author} {\bibfnamefont
  {P.}~\bibnamefont {Heu}}, \bibinfo {author} {\bibfnamefont {T.}~\bibnamefont
  {Corbitt}}, \ and\ \bibinfo {author} {\bibfnamefont {N.}~\bibnamefont
  {Mavalvala}},\ }\href@noop {} {\bibfield  {journal} {\bibinfo  {journal}
  {arXiv preprint arXiv:1812.09942}\ } (\bibinfo {year} {2018})}\BibitemShut
  {NoStop}%
\bibitem [{\citenamefont {Vitali}\ \emph {et~al.}(2007)\citenamefont {Vitali},
  \citenamefont {Gigan}, \citenamefont {Ferreira}, \citenamefont {B\"ohm},
  \citenamefont {Tombesi}, \citenamefont {Guerreiro}, \citenamefont {Vedral},
  \citenamefont {Zeilinger},\ and\ \citenamefont
  {Aspelmeyer}}]{PhysRevLett.98.030405}%
  \BibitemOpen
  \bibfield  {author} {\bibinfo {author} {\bibfnamefont {D.}~\bibnamefont
  {Vitali}}, \bibinfo {author} {\bibfnamefont {S.}~\bibnamefont {Gigan}},
  \bibinfo {author} {\bibfnamefont {A.}~\bibnamefont {Ferreira}}, \bibinfo
  {author} {\bibfnamefont {H.~R.}\ \bibnamefont {B\"ohm}}, \bibinfo {author}
  {\bibfnamefont {P.}~\bibnamefont {Tombesi}}, \bibinfo {author} {\bibfnamefont
  {A.}~\bibnamefont {Guerreiro}}, \bibinfo {author} {\bibfnamefont
  {V.}~\bibnamefont {Vedral}}, \bibinfo {author} {\bibfnamefont
  {A.}~\bibnamefont {Zeilinger}}, \ and\ \bibinfo {author} {\bibfnamefont
  {M.}~\bibnamefont {Aspelmeyer}},\ }\href {\doibase
  10.1103/PhysRevLett.98.030405} {\bibfield  {journal} {\bibinfo  {journal}
  {Phys. Rev. Lett.}\ }\textbf {\bibinfo {volume} {98}},\ \bibinfo {pages}
  {030405} (\bibinfo {year} {2007})}\BibitemShut {NoStop}%
\bibitem [{\citenamefont {Wipf}\ \emph {et~al.}(2008)\citenamefont {Wipf},
  \citenamefont {Corbitt}, \citenamefont {Chen},\ and\ \citenamefont
  {Mavalvala}}]{Wipf_2008}%
  \BibitemOpen
  \bibfield  {author} {\bibinfo {author} {\bibfnamefont {C.}~\bibnamefont
  {Wipf}}, \bibinfo {author} {\bibfnamefont {T.}~\bibnamefont {Corbitt}},
  \bibinfo {author} {\bibfnamefont {Y.}~\bibnamefont {Chen}}, \ and\ \bibinfo
  {author} {\bibfnamefont {N.}~\bibnamefont {Mavalvala}},\ }\href {\doibase
  10.1088/1367-2630/10/9/095017} {\bibfield  {journal} {\bibinfo  {journal}
  {New Journal of Physics}\ }\textbf {\bibinfo {volume} {10}},\ \bibinfo
  {pages} {095017} (\bibinfo {year} {2008})}\BibitemShut {NoStop}%
\bibitem [{\citenamefont {Wipf}(2013)}]{wipf_2013}%
  \BibitemOpen
  \bibfield  {author} {\bibinfo {author} {\bibfnamefont {C.}~\bibnamefont
  {Wipf}},\ }\emph {\bibinfo {title} {Toward quantum opto-mechanics in a
  gram-scale suspended mirror interferometer}},\ \href
  {http://hdl.handle.net/1721.1/84182} {Ph.D. thesis},\ \bibinfo  {school}
  {Massachusetts Institute of Technology} (\bibinfo {year} {2013})\BibitemShut
  {NoStop}%
\bibitem [{\citenamefont {Wang}\ and\ \citenamefont
  {Clerk}(2013)}]{PhysRevLett.110.253601}%
  \BibitemOpen
  \bibfield  {author} {\bibinfo {author} {\bibfnamefont {Y.-D.}\ \bibnamefont
  {Wang}}\ and\ \bibinfo {author} {\bibfnamefont {A.~A.}\ \bibnamefont
  {Clerk}},\ }\href {\doibase 10.1103/PhysRevLett.110.253601} {\bibfield
  {journal} {\bibinfo  {journal} {Phys. Rev. Lett.}\ }\textbf {\bibinfo
  {volume} {110}},\ \bibinfo {pages} {253601} (\bibinfo {year}
  {2013})}\BibitemShut {NoStop}%
\bibitem [{\citenamefont {Barzanjeh}\ \emph {et~al.}(2019)\citenamefont
  {Barzanjeh}, \citenamefont {Redchenko}, \citenamefont {Peruzzo},
  \citenamefont {Wulf}, \citenamefont {Lewis}, \citenamefont {Arnold},\ and\
  \citenamefont {Fink}}]{Barzanjeh2019}%
  \BibitemOpen
  \bibfield  {author} {\bibinfo {author} {\bibfnamefont {S.}~\bibnamefont
  {Barzanjeh}}, \bibinfo {author} {\bibfnamefont {E.~S.}\ \bibnamefont
  {Redchenko}}, \bibinfo {author} {\bibfnamefont {M.}~\bibnamefont {Peruzzo}},
  \bibinfo {author} {\bibfnamefont {M.}~\bibnamefont {Wulf}}, \bibinfo {author}
  {\bibfnamefont {D.~P.}\ \bibnamefont {Lewis}}, \bibinfo {author}
  {\bibfnamefont {G.}~\bibnamefont {Arnold}}, \ and\ \bibinfo {author}
  {\bibfnamefont {J.~M.}\ \bibnamefont {Fink}},\ }\href {\doibase
  10.1038/s41586-019-1320-2} {\bibfield  {journal} {\bibinfo  {journal}
  {Nature}\ }\textbf {\bibinfo {volume} {570}},\ \bibinfo {pages} {480}
  (\bibinfo {year} {2019})}\BibitemShut {NoStop}%
\bibitem [{\citenamefont {Bienfait}\ \emph {et~al.}(2019)\citenamefont
  {Bienfait}, \citenamefont {Satzinger}, \citenamefont {Zhong}, \citenamefont
  {Chang}, \citenamefont {Chou}, \citenamefont {Conner}, \citenamefont {Dumur},
  \citenamefont {Grebel}, \citenamefont {Peairs}, \citenamefont {Povey},\ and\
  \citenamefont {Cleland}}]{Bienfait368}%
  \BibitemOpen
  \bibfield  {author} {\bibinfo {author} {\bibfnamefont {A.}~\bibnamefont
  {Bienfait}}, \bibinfo {author} {\bibfnamefont {K.~J.}\ \bibnamefont
  {Satzinger}}, \bibinfo {author} {\bibfnamefont {Y.~P.}\ \bibnamefont
  {Zhong}}, \bibinfo {author} {\bibfnamefont {H.-S.}\ \bibnamefont {Chang}},
  \bibinfo {author} {\bibfnamefont {M.-H.}\ \bibnamefont {Chou}}, \bibinfo
  {author} {\bibfnamefont {C.~R.}\ \bibnamefont {Conner}}, \bibinfo {author}
  {\bibfnamefont {{\'E}.}~\bibnamefont {Dumur}}, \bibinfo {author}
  {\bibfnamefont {J.}~\bibnamefont {Grebel}}, \bibinfo {author} {\bibfnamefont
  {G.~A.}\ \bibnamefont {Peairs}}, \bibinfo {author} {\bibfnamefont {R.~G.}\
  \bibnamefont {Povey}}, \ and\ \bibinfo {author} {\bibfnamefont {A.~N.}\
  \bibnamefont {Cleland}},\ }\href {\doibase 10.1126/science.aaw8415}
  {\bibfield  {journal} {\bibinfo  {journal} {Science}\ }\textbf {\bibinfo
  {volume} {364}},\ \bibinfo {pages} {368} (\bibinfo {year} {2019})},\ \Eprint
  {http://arxiv.org/abs/https://science.sciencemag.org/content/364/6438/368.full.pdf}
  {https://science.sciencemag.org/content/364/6438/368.full.pdf} \BibitemShut
  {NoStop}%
\bibitem [{\citenamefont {Chen}\ \emph {et~al.}(2020)\citenamefont {Chen},
  \citenamefont {Rossi}, \citenamefont {Mason},\ and\ \citenamefont
  {Schliesser}}]{Chen2020}%
  \BibitemOpen
  \bibfield  {author} {\bibinfo {author} {\bibfnamefont {J.}~\bibnamefont
  {Chen}}, \bibinfo {author} {\bibfnamefont {M.}~\bibnamefont {Rossi}},
  \bibinfo {author} {\bibfnamefont {D.}~\bibnamefont {Mason}}, \ and\ \bibinfo
  {author} {\bibfnamefont {A.}~\bibnamefont {Schliesser}},\ }\href {\doibase
  10.1038/s41467-020-14768-1} {\bibfield  {journal} {\bibinfo  {journal}
  {Nature Communications}\ }\textbf {\bibinfo {volume} {11}},\ \bibinfo {pages}
  {943} (\bibinfo {year} {2020})}\BibitemShut {NoStop}%
\bibitem [{\citenamefont {Corbitt}\ \emph {et~al.}(2007)\citenamefont
  {Corbitt}, \citenamefont {Chen}, \citenamefont {Innerhofer}, \citenamefont
  {M\"uller-Ebhardt}, \citenamefont {Ottaway}, \citenamefont {Rehbein},
  \citenamefont {Sigg}, \citenamefont {Whitcomb}, \citenamefont {Wipf},\ and\
  \citenamefont {Mavalvala}}]{PhysRevLett.98.150802}%
  \BibitemOpen
  \bibfield  {author} {\bibinfo {author} {\bibfnamefont {T.}~\bibnamefont
  {Corbitt}}, \bibinfo {author} {\bibfnamefont {Y.}~\bibnamefont {Chen}},
  \bibinfo {author} {\bibfnamefont {E.}~\bibnamefont {Innerhofer}}, \bibinfo
  {author} {\bibfnamefont {H.}~\bibnamefont {M\"uller-Ebhardt}}, \bibinfo
  {author} {\bibfnamefont {D.}~\bibnamefont {Ottaway}}, \bibinfo {author}
  {\bibfnamefont {H.}~\bibnamefont {Rehbein}}, \bibinfo {author} {\bibfnamefont
  {D.}~\bibnamefont {Sigg}}, \bibinfo {author} {\bibfnamefont {S.}~\bibnamefont
  {Whitcomb}}, \bibinfo {author} {\bibfnamefont {C.}~\bibnamefont {Wipf}}, \
  and\ \bibinfo {author} {\bibfnamefont {N.}~\bibnamefont {Mavalvala}},\ }\href
  {\doibase 10.1103/PhysRevLett.98.150802} {\bibfield  {journal} {\bibinfo
  {journal} {Phys. Rev. Lett.}\ }\textbf {\bibinfo {volume} {98}},\ \bibinfo
  {pages} {150802} (\bibinfo {year} {2007})}\BibitemShut {NoStop}%
\bibitem [{\citenamefont {Sharifi}\ \emph {et~al.}(2019)\citenamefont
  {Sharifi}, \citenamefont {Banadaki}, \citenamefont {Cullen}, \citenamefont
  {Veronis}, \citenamefont {Dowling},\ and\ \citenamefont
  {Corbitt}}]{sharifi2019design}%
  \BibitemOpen
  \bibfield  {author} {\bibinfo {author} {\bibfnamefont {S.}~\bibnamefont
  {Sharifi}}, \bibinfo {author} {\bibfnamefont {Y.}~\bibnamefont {Banadaki}},
  \bibinfo {author} {\bibfnamefont {T.}~\bibnamefont {Cullen}}, \bibinfo
  {author} {\bibfnamefont {G.}~\bibnamefont {Veronis}}, \bibinfo {author}
  {\bibfnamefont {J.}~\bibnamefont {Dowling}}, \ and\ \bibinfo {author}
  {\bibfnamefont {T.}~\bibnamefont {Corbitt}},\ }\href@noop {} {\enquote
  {\bibinfo {title} {Design of microresonators to minimize thermal noise below
  the standard quantum limit},}\ } (\bibinfo {year} {2019}),\ \Eprint
  {http://arxiv.org/abs/1911.02200} {arXiv:1911.02200 [quant-ph]} \BibitemShut
  {NoStop}%
\bibitem [{\citenamefont {Cripe}(2018)}]{cripe_2018}%
  \BibitemOpen
  \bibfield  {author} {\bibinfo {author} {\bibfnamefont {J.~D.}\ \bibnamefont
  {Cripe}},\ }\emph {\bibinfo {title} {Broadband measurement and reduction of
  quantum radiation pressure noise in the audio band}},\ \href
  {https://digitalcommons.lsu.edu/gradschool_dissertations/4653} {Ph.D.
  thesis},\ \bibinfo  {school} {LSU Doctoral Dissertations} (\bibinfo {year}
  {2018})\BibitemShut {NoStop}%
\bibitem [{\citenamefont {Vidal}\ and\ \citenamefont
  {Werner}(2002)}]{PhysRevA.65.032314}%
  \BibitemOpen
  \bibfield  {author} {\bibinfo {author} {\bibfnamefont {G.}~\bibnamefont
  {Vidal}}\ and\ \bibinfo {author} {\bibfnamefont {R.~F.}\ \bibnamefont
  {Werner}},\ }\href {\doibase 10.1103/PhysRevA.65.032314} {\bibfield
  {journal} {\bibinfo  {journal} {Phys. Rev. A}\ }\textbf {\bibinfo {volume}
  {65}},\ \bibinfo {pages} {032314} (\bibinfo {year} {2002})}\BibitemShut
  {NoStop}%
\bibitem [{\citenamefont {Plenio}(2005)}]{PhysRevLett.95.090503}%
  \BibitemOpen
  \bibfield  {author} {\bibinfo {author} {\bibfnamefont {M.~B.}\ \bibnamefont
  {Plenio}},\ }\href {\doibase 10.1103/PhysRevLett.95.090503} {\bibfield
  {journal} {\bibinfo  {journal} {Phys. Rev. Lett.}\ }\textbf {\bibinfo
  {volume} {95}},\ \bibinfo {pages} {090503} (\bibinfo {year}
  {2005})}\BibitemShut {NoStop}%
\bibitem [{\citenamefont {Audenaert}\ \emph {et~al.}(2003)\citenamefont
  {Audenaert}, \citenamefont {Plenio},\ and\ \citenamefont
  {Eisert}}]{PhysRevLett.90.027901}%
  \BibitemOpen
  \bibfield  {author} {\bibinfo {author} {\bibfnamefont {K.}~\bibnamefont
  {Audenaert}}, \bibinfo {author} {\bibfnamefont {M.~B.}\ \bibnamefont
  {Plenio}}, \ and\ \bibinfo {author} {\bibfnamefont {J.}~\bibnamefont
  {Eisert}},\ }\href {\doibase 10.1103/PhysRevLett.90.027901} {\bibfield
  {journal} {\bibinfo  {journal} {Phys. Rev. Lett.}\ }\textbf {\bibinfo
  {volume} {90}},\ \bibinfo {pages} {027901} (\bibinfo {year}
  {2003})}\BibitemShut {NoStop}%
\bibitem [{\citenamefont {Wang}\ and\ \citenamefont
  {Wilde}(2018)}]{wang2018exact}%
  \BibitemOpen
  \bibfield  {author} {\bibinfo {author} {\bibfnamefont {X.}~\bibnamefont
  {Wang}}\ and\ \bibinfo {author} {\bibfnamefont {M.~M.}\ \bibnamefont
  {Wilde}},\ }\href@noop {} {\enquote {\bibinfo {title} {Exact entanglement
  cost of quantum states and channels under ppt-preserving operations},}\ }
  (\bibinfo {year} {2018}),\ \Eprint {http://arxiv.org/abs/1809.09592}
  {arXiv:1809.09592 [quant-ph]} \BibitemShut {NoStop}%
\bibitem [{\citenamefont {Wang}\ and\ \citenamefont {Wilde}(2020)}]{WangWilde}%
  \BibitemOpen
  \bibfield  {author} {\bibinfo {author} {\bibfnamefont {X.}~\bibnamefont
  {Wang}}\ and\ \bibinfo {author} {\bibfnamefont {M.~M.}\ \bibnamefont
  {Wilde}},\ }\href {\doibase 10.1103/PhysRevLett.125.} {\bibfield  {journal}
  {\bibinfo  {journal} {Phys. Rev. Lett.}\ }\textbf {\bibinfo {volume} {125}}
  (\bibinfo {year} {2020}),\ 10.1103/PhysRevLett.125.}\BibitemShut {Stop}%
\bibitem [{\citenamefont {Serafini}(2017)}]{serafini_2017}%
  \BibitemOpen
  \bibfield  {author} {\bibinfo {author} {\bibfnamefont {A.}~\bibnamefont
  {Serafini}},\ }\href@noop {} {\emph {\bibinfo {title} {Quantum Continuous
  Variables: A Primer of Theoretical Methods}}},\ \bibinfo {edition} {1st}\
  ed.,\ Vol.~\bibinfo {volume} {1}\ (\bibinfo  {publisher} {CRC Press},\
  \bibinfo {year} {2017})\BibitemShut {NoStop}%
\bibitem [{\citenamefont {Corbitt}\ \emph {et~al.}(2005)\citenamefont
  {Corbitt}, \citenamefont {Chen},\ and\ \citenamefont
  {Mavalvala}}]{PhysRevA.72.013818}%
  \BibitemOpen
  \bibfield  {author} {\bibinfo {author} {\bibfnamefont {T.}~\bibnamefont
  {Corbitt}}, \bibinfo {author} {\bibfnamefont {Y.}~\bibnamefont {Chen}}, \
  and\ \bibinfo {author} {\bibfnamefont {N.}~\bibnamefont {Mavalvala}},\ }\href
  {\doibase 10.1103/PhysRevA.72.013818} {\bibfield  {journal} {\bibinfo
  {journal} {Phys. Rev. A}\ }\textbf {\bibinfo {volume} {72}},\ \bibinfo
  {pages} {013818} (\bibinfo {year} {2005})}\BibitemShut {NoStop}%
\bibitem [{\citenamefont {Caves}\ and\ \citenamefont
  {Schumaker}(1985)}]{PhysRevA.31.3068}%
  \BibitemOpen
  \bibfield  {author} {\bibinfo {author} {\bibfnamefont {C.~M.}\ \bibnamefont
  {Caves}}\ and\ \bibinfo {author} {\bibfnamefont {B.~L.}\ \bibnamefont
  {Schumaker}},\ }\href {\doibase 10.1103/PhysRevA.31.3068} {\bibfield
  {journal} {\bibinfo  {journal} {Phys. Rev. A}\ }\textbf {\bibinfo {volume}
  {31}},\ \bibinfo {pages} {3068} (\bibinfo {year} {1985})}\BibitemShut
  {NoStop}%
\bibitem [{\citenamefont {Rains}(1999)}]{PhysRevA.60.179}%
  \BibitemOpen
  \bibfield  {author} {\bibinfo {author} {\bibfnamefont {E.~M.}\ \bibnamefont
  {Rains}},\ }\href {\doibase 10.1103/PhysRevA.60.179} {\bibfield  {journal}
  {\bibinfo  {journal} {Phys. Rev. A}\ }\textbf {\bibinfo {volume} {60}},\
  \bibinfo {pages} {179} (\bibinfo {year} {1999})}\BibitemShut {NoStop}%
\bibitem [{\citenamefont {Bäuml}\ \emph {et~al.}(2019)\citenamefont {Bäuml},
  \citenamefont {Das}, \citenamefont {Wang},\ and\ \citenamefont
  {Wilde}}]{buml2019resource}%
  \BibitemOpen
  \bibfield  {author} {\bibinfo {author} {\bibfnamefont {S.}~\bibnamefont
  {Bäuml}}, \bibinfo {author} {\bibfnamefont {S.}~\bibnamefont {Das}},
  \bibinfo {author} {\bibfnamefont {X.}~\bibnamefont {Wang}}, \ and\ \bibinfo
  {author} {\bibfnamefont {M.~M.}\ \bibnamefont {Wilde}},\ }\href@noop {}
  {\enquote {\bibinfo {title} {Resource theory of entanglement for bipartite
  quantum channels},}\ } (\bibinfo {year} {2019}),\ \Eprint
  {http://arxiv.org/abs/1907.04181} {arXiv:1907.04181 [quant-ph]} \BibitemShut
  {NoStop}%
\bibitem [{\citenamefont {Duan}\ \emph {et~al.}(2000)\citenamefont {Duan},
  \citenamefont {Giedke}, \citenamefont {Cirac},\ and\ \citenamefont
  {Zoller}}]{PhysRevLett.84.2722}%
  \BibitemOpen
  \bibfield  {author} {\bibinfo {author} {\bibfnamefont {L.-M.}\ \bibnamefont
  {Duan}}, \bibinfo {author} {\bibfnamefont {G.}~\bibnamefont {Giedke}},
  \bibinfo {author} {\bibfnamefont {J.~I.}\ \bibnamefont {Cirac}}, \ and\
  \bibinfo {author} {\bibfnamefont {P.}~\bibnamefont {Zoller}},\ }\href
  {\doibase 10.1103/PhysRevLett.84.2722} {\bibfield  {journal} {\bibinfo
  {journal} {Phys. Rev. Lett.}\ }\textbf {\bibinfo {volume} {84}},\ \bibinfo
  {pages} {2722} (\bibinfo {year} {2000})}\BibitemShut {NoStop}%
\end{thebibliography}%

\end{document}